\documentclass[11pt]{article}
\usepackage{moriond,epsfig}

\def\be{\begin{equation}}
\def\ee{\end{equation}}
\def\bea{\begin{eqnarray}}
\def\eea{\end{eqnarray}}

\begin{document}
\vspace*{4cm}
\title{
$K_{e3}$ AND $\pi_{e3}$ 
DECAYS: RADIATIVE CORRECTIONS AND CKM
UNITARITY
}

\author{ V. CIRIGLIANO }

\address{
Departament de F\'{\i}sica Te\`orica, IFIC, CSIC ---
Universitat de Val\`encia \\
Apt. Correus 2085, E-46071  Val\`encia, Spain 
}

\maketitle\abstracts{
We discuss radiative corrections to $K_{e3}$ and
$\pi_{e3}$ within chiral perturbation theory with virtual photons and
leptons.  We then present the extraction of the CKM elements $V_{us}$
and $V_{ud}$ using the presently available experimental input. 
Finally we discuss the test of CKM unitarity based
on the present independent knowledge of $V_{ud}$ and $V_{us}$, and
describe the prospects of improvement with the upcoming new
B.R. measurements. }

\section{Motivation}

$V_{us}$ and $V_{ud}$ are, at present, the most precisely known
elements of the CKM matrix, with a fractional uncertainty 
at the level of $\sim 1\%$ and $\sim 0.1\%$ respectively. 
This high-precision information is extracted from the semileptonic
transitions $s\rightarrow u$ and $d\rightarrow u$, occurring in
low-energy hadronic environments.  In particular, the best
determination of $|V_{us}|$ is obtained from $K \to \pi \ell \nu$
decays ($K_{\ell3}$), whereas the two most stringent constraints on
$|V_{ud}|$ are obtained from super-allowed Fermi transitions (SFT), and
from the neutron beta decay~\cite{ckm03}. 
The high accuracy of these constraints is due to two key features:
the non-renormalization of the vector current at zero momentum transfer
in the $SU(N)$ limit (CVC), 
and the Ademollo Gatto theorem \cite{ag}. 
The latter implies that corrections to 
the relevant hadronic form factors at zero momentum transfer are of second 
order in the $SU(N)$-breaking parameters (N=2,3).  

The present level of accuracy on $|V_{ud}|$ and $|V_{us}|$ is such
that the contribution of $|V_{ub}|$ to the unitarity relation
\begin{equation}
U_{uu}=|V_{ud}|^2+|V_{us}|^2+|V_{ub}|^2 =1 
\label{eq:unitarity}
\end{equation}
can be safely neglected, and the uncertainty of the first two terms is
comparable.  In other words, $|V_{ud}|$ and $|V_{us}|$ lead to two
independent determinations of the Cabibbo angle both at the 1\% level.
Plugging in the PDG numbers~\cite{pdg2002}, one finds a discrepancy 
between the two determinations at the two-sigma level.  
An opportunity to clarify this long-standing puzzle is offered us by
the ongoing experimental efforts which will provide new measurements
of $\pi_{e3}$~\cite{PSI} and $K_{\ell 3}$ branching
ratios~\cite{E865,KLOE,NA48}.  Precise data on $\pi_{e3}$
(B.R. $\sim 10^{-8}$) will allow one to extract $V_{ud}$ from a
theoretically very clean environment, while improved $K_{\ell 3}$
B.R.s  will allow one to update the extraction of $V_{us}$. 
In both cases, however, in order to fully take advantage of new data,
an update of the theoretical analysis is in order. In particular, one
needs to have good control over the $SU(2)$ and $SU(3)$ breaking
effects, as well as on radiative corrections.
Chiral Perturbation Theory provides a modern unified framework for
studying both problems. 

\section{Semileptonic decays of light mesons}

\subsection{Theoretical Framework}

Weak decays of hadrons  are best treated within an effective theory
approach. In the case at hand, 
one starts by assuming that the Standard Model is the appropriate theory
at the scale $\mu \sim 100$ GeV.  At scales $\mu \sim M_W$ one
integrates out the heavy degrees of freedom  and matches
the SM to the Fermi Theory of weak interaction, enhanced 
by QED and QCD, with five active quark flavors. 
The effective lagrangian governing semileptonic 
transitions assumes the form:
\begin{eqnarray}
{\cal L}_{\mathrm{eff}} &=& - \left. \displaystyle \frac{{G_\mu}}{\sqrt{2}} 
\sqrt{S_{\mbox{\tiny ew}} (\nu)} \times {\cal O}_{JJ} (\nu)  
\right.  \ ,   \nonumber \\
{\cal O}_{JJ} &=&  \bar{l}_k \, \gamma_\mu (1 - \gamma_5) \, \nu_{k} 
\, \times \, 
\bar{u}_i  \, \gamma^\mu (1 - \gamma_5)\,  {\bf V}_{ij} d_j  \ . 
\end{eqnarray}
Here $i,j,k$ are flavor indices (running over active flavors), 
and ${\bf V}$ is the CKM matrix. 
$G_{\mu}$ is the Fermi constant governing purely leptonic charged current 
processes (measured in the muon decay).   
Finally, $\nu$ indicates the renormalization scale. 

The effective coupling is known up to non-logarithmic terms of
order $\alpha \alpha_s$, and is given by~\cite{shortdist}:
\begin{equation}
S_{\rm ew} (\mu) = 1 + \frac{2 \alpha}{\pi} \left( 
1 - \frac{\alpha_s}{4 \pi} \right) 
\log \frac{M_Z}{\nu}  + O (\frac{\alpha \alpha_s}{\pi^2})  
\end{equation}
After re-summing the leading logarithms $O[ (\alpha \log M_Z)^n, \alpha
(\alpha_s \log M_Z)^n]$ via renormalization group, 
down to the  renormalization point $\nu=M_\rho$, one finds
(taking into account fermion thresholds):
\begin{equation}
S_{\rm ew} (M_{\rho}) = 1.0230 \pm 0.0003  \ . 
\end{equation}

The second part of the problem consists in calculating matrix
elements of the type $\langle f | {\cal O}_{JJ} | i \rangle$. 
In our case, the relevant matrix element has the form:
\begin{equation}
\langle \pi \ell \nu | {\cal O}_{JJ} | P \rangle \sim 
J_l  J_{h} \left[ 1 + 
O \left( \alpha \log \frac{M_P}{m_e}, \alpha c_n \right)
\right] \ , 
\end{equation}
where $J_l$ is the leptonic current and $J_h$ is the hadronic current. 
The radiative corrections will in general involve large long-distance 
logs as well as non-logarithmic pieces ($\alpha c_n$). 
An accurate calculation of semileptonic decays requires control over
$J_h$ (QCD dynamics at low energy), and the potentially large 
long-distance radiative corrections. 

The appropriate framework to address both issues is provided by yet
another effective theory, valid in the energy range below the first
QCD resonance masses: Chiral Perturbation Theory (ChPT)~\cite{chpt}.  Active
degrees of freedom in ChPT are the eight Goldostone modes ($\pi, K,
\eta$) associated with SSB of chiral symmetry in QCD, as well as light
leptons and photons.  The effective lagrangian of ChPT is constructed
according to symmetry principles, taking into account the SSB pattern
of chiral symmetry as well as  explicit symmetry breaking, due to 
quark masses.  The effective action is
organized by chiral power counting as an expansion in the derivatives
(momenta) of Goldstone modes (counted as $O(p)$) and the quark masses
(counted as $O(p^2)$ in the standard approach).  At a given order in
$p/\Lambda_{\chi}$ (with $\Lambda_{\chi} \sim 4 \pi F_{\pi} $) 
a finite number of local couplings encodes the short distance
physics.  Even though a full matching of QCD to ChPT is not available,
model-independent phenomenology is possible in several cases of
interest.  In this work we need the extension of ChPT
developed in Ref.~\cite{lept}, appropriate to deal with virtual
photons and leptons.

\subsection{$P_{\ell 3}$ decays: matrix elements and form factors}

Neglecting for the moment radiative corrections, 
the invariant amplitude 
for  the decay $ P (p_P) \longrightarrow  \pi (p_\pi) \ \ell^+ (p_\ell) \  
\nu_\ell (p_\nu) $ 
factorizes in the product of leptonic 
and hadronic currents, the latter being decomposed in terms of 
hadronic form factors. Explicitly one has 
\begin{equation}
{\cal M} =  \frac{G_{\mu}}{\sqrt{2}} \,    V^{*}_{u f}   \,  L^{\mu} \, 
\underbrace{ { C_P} \, \bigg[ f_{+}^{P} (t) \, (p_P + p_\pi)_{\mu} 
+  f_{-}^{P} (t) \, (p_P - p_\pi)_{\mu} \bigg]}_{
\langle \pi (p_\pi) | {   \bar{u} \gamma_{\mu}  f} | P (p_P) \rangle }
\ , 
\end{equation}
with 
\begin{equation}
\begin{array}{lll}
L^\mu & = & \bar{u} (p_\nu)  \gamma^\mu  (1 - \gamma_5)  v (p_\ell) \\ 
t & = & (p_P - p_\pi)^2 
\end{array} 
\qquad C_P = (1,1/\sqrt{2},\sqrt{2}) \  \mbox{for} \ (K^0,K^{+},\pi^+)
\end{equation}
We recall that the dependence of $|{\cal M}|^2$ on $f_-$ 
is proportional to $(m_\ell/M_P)^2$, thus $f_-$ is completely irrelevant 
for the electron modes ($K_{e3}, \pi_{e3}$). 
Moreover, it is customary to trade 
$f_- (t)$ for the scalar form factor $f_0 (t)$, defined  as: 
\begin{equation}
f_0 (t) = f_+ (t) \ + \ \frac{t}{M_K^2 - M_\pi^2} \, f_- (t) \ . 
\end{equation}
The momentum dependence of the form factors, relevant for the
integral over the phase space, is often described in terms of 
one or at most two parameters (slope and curvature at $t=0$), 
\begin{equation}
f_{+,0} (t) = {  f_{+} (0) }  \, \left( 1 + {  \lambda_{+,0} } 
\ \frac{t}{M_\pi^2} +  {  \lambda_{+,0}'  } \frac{t^2}{M_\pi^4}  
+ \dots \right) \ . 
\end{equation}
In this approximation the phase space integral depends explicitly 
only on the slope (and curvature) parameters, and we use for it the 
generic notation $I_P (\lambda)$. 
With this notation, the partial width associated to  
$P \rightarrow \pi \ell \nu$ is given by:
\begin{equation}
\Gamma_P = 
|V_{uf}|^2 \
\cdot | {  f_{+}^{P} (0)} |^2 \cdot  
{\cal N}_P   \cdot  I_P ({  \lambda})   \ ; 
\qquad {\cal N}_P =  C_P^2 \  \frac{G_{\mu}^2 \  M_P^5}{192 \pi^3} \ . 
\label{decayrate}
\end{equation}
The steps necessary to extract $|V_{uf}| (f=d,s)$  from the 
measured $P_{\ell 3}$ decay rates are:
(1) theoretical evaluation of $f_+(0)$, i.e. the $SU(2)$ and 
$SU(3)$ corrections;
(2) measurement (or, if not available, theoretical evaluation)
of the momentum dependence of $f_{+,0} (t)$ (form factor slope 
and curvature parameters); 
(3) theoretical treatment of photonic radiative corrections. 
The short distance contribution is given by the factor $S_{\rm ew}$ 
and amounts to using a different effective Fermi coupling, 
while the long-distance corrections will be discussed in the 
next subsection  [note that Eq.\ref{decayrate}
is not yet general enough to account for these effects, see below].

\subsection{Long distance radiative corrections} 

A recent analysis of this issue has been performed in Ref.~\cite{CKNRT}, 
to which we refer for details. 
Long distance radiative corrections involve the effect of (1) loops 
with virtual photons and (2) emission of real photons:  
\begin{enumerate}
\item  
Photon loops modify the very  structure of the amplitude, 
breaking the factorized form.  
The form factors now depend on another kinematical variable 
$$f_{\pm} (t) \rightarrow  f_{\pm}(t,v) \simeq \left[ 1 + 
\frac{\alpha}{4 \pi} \Gamma_c (v, \lambda_{IR}) \right] 
\, \tilde{f}_{\pm} (t) \ , $$
where $v=(p_P - p_l)^2$ in $P^+$ decays and 
$v=(p_\pi + p_l)^2$ in $P^0$ decays.
The function $\Gamma_c (v, \lambda_{IR})$ encodes universal long
distance corrections, independent of strong dynamics, and 
is infrared divergent.  Since the dependence on the second
kinematical variable can be factored out (to a very good
approximation), the notion of effective form factor $\tilde{f}_{\pm}
(t) $ survives and proves useful in the subsequent analysis.
$\tilde{f}_{\pm} (0)$ contains pure QCD effects as well as new local
contributions of order $\alpha$.  These, together with the chiral
logarithms, are truly structure dependent corrections, which can be
described in a model independent way within the ChPT approach. 

\item
One has to consider how radiation 
of real photons affects the various observables (e.g. Dalitz Plot 
density, spectra, branching ratios). For the purpose of extracting 
$|V_{uf}|$, we need to assess the effect of real photon emission 
on the partial widths. As is well known, a given experiment measures 
an inclusive sum of the parent mode and radiative modes:
$$ d \Gamma_{\rm obs} = d \Gamma (P_{\ell 3}) \ + \ 
 d \Gamma (P_{\ell 3 \gamma}) \ + \ \cdots $$ 
From the theoretical point of view, only such an inclusive sum 
is free of infrared singularities. 
At the precision we aim to work at, a meaningful comparison of theory
and experiment can be done only once a clear definition of the inclusive
observable is given~\cite{ginsberg}. 

\end{enumerate}

In summary, all the universal QED effects, due to both virtual photons
[$\Gamma_c (v,\lambda_{IR})$] and real photons [$d \Gamma (K_{\ell 3
\gamma})$], can be combined to produce a correction to the phase space
factor in the expression for the decay width:
$$ 
I_P (\lambda) \longrightarrow 
I_{P} ( \lambda, \alpha) = 
I_{P} (\lambda, 0) \, \bigg[ 
1 + \Delta I_{P} (\lambda) \bigg] \ . 
$$
This term comes, in principle, with no theoretical uncertainty.  The
structure dependent electromagnetic corrections, as well as 
 $SU(3)$ breaking corrections  are in the form factor
$\tilde{f}_{+} (t)$, where all of the theoretical uncertainty
concentrates. At zero momentum transfer, this can be conveniently written as
$$  
\tilde{f}_{+} (0) = f_{+} (0) \, ( 1 + \delta_{SU(2)} + \delta_{EM} )
$$ 
Explicit expressions for  $\tilde{f}_{\pm} (t)$ at NLO in ChPT, including 
the SU(2) and EM corrections,  can be found in Refs.~\cite{CKNRT,cknp02}.


\section{$V_{ud}$ from $\pi_{e3}$ decay}

Theoretically this process is very attractive because it 
shares the advantages of both Fermi transitions (pure vector transition, 
no axial-vector contribution) and neutron $\beta$-decay (no nuclear structure
dependent radiative corrections).
The difficulty here lies in the extremely small branching  ratio, 
of order $10^{-8}$.
With the notations introduced in the previous sections, $V_{ud}$ is 
related to the partial width by the following relation: 
\begin{equation}
|V_{ud}| = \left[ 
\displaystyle\frac{\Gamma_{\pi_{e 3[\gamma]}}}{ {\cal N}_\pi \,
 I_{\pi} (\lambda, 0) } \right]^{1/2} \, 
\left[ \displaystyle\frac{1}{S_{\rm ew} (M_\rho)} \right]^{1/2} \, 
\displaystyle\frac{1}{1 + \delta^{\pi}_{SU(2)} + 
 \delta^{\pi}_{\rm EM} + \frac{1}{2}
\Delta I_{\pi} (\lambda)}  
\label{masterVud} 
\end{equation}
Explicit calculation of the radiative corrections and $SU(2)$ breaking 
leads to~\cite{cknp02}: 
\begin{equation}
\delta^{\pi}_{SU(2)} \sim 10^{-5} 
\ , \qquad  
\delta^{\pi}_{e^2 p^2} = (0.46 \pm 0.05) \% \ , \qquad 
 \Delta I_{\pi} (\lambda) = 0.1 \%  \ ,
\end{equation}
with a total effect of radiative corrections consistent 
with previous estimates \cite{pibeta}. \\
The present experimental precision for the branching ratio of the pionic 
beta decay cannot compete yet with the very small theoretical 
uncertainties of SFT and neutron beta decay: 
using the latest PDG value \cite{pdg2002}
${\rm BR} = (1.025 \pm 0.034) \times 10^{-8}$, we find
\begin{equation}|
V_{ud}| = 0.9675 \pm 0.0160_{\rm exp} \pm 0.0005_{\rm th}
         = 0.9675 \pm 0.0161 .
\end{equation}
However, a substantial improvement of the experimental accuracy is to be 
expected in the near future. Inserting the present preliminary result
obtained by the PIBETA Collaboration \cite{PSI},
$ {\rm BR} = (1.044 \pm 0.007 ({\rm stat.}) \pm 0.009 ({\rm syst.})) \times 
10^{-8}$, we find
\begin{equation}
|V_{ud}| = 0.9765 \pm 0.0056_{\rm exp} \pm 0.0005_{\rm th}
= 0.9765 \pm 0.0056~,
\end{equation}
where the error should be reduced by about a factor of 3 at the 
end of the experiment.

\section{$V_{us}$ from $K_{e3}$ decays}


\begin{table}[t]
\caption{Summary of isospin-breaking factors.
\label{tab:iso-brk}}
\vspace{0.4cm}
\centering
\begin{tabular}{|c|c|c|c|}
\hline
  & $\delta_{SU(2)} (\%)$ & $\delta_{\rm EM} (\%)$ &
$\Delta I (\lambda)  (\%)$
\\
\hline
$K^{+}_{e3}$ & 2.4 $\pm$ 0.2  & 0.32 $\pm$ 0.16 & -1.27 
\\
$K^{0}_{e 3}$ &    0    & 0.46 $\pm$ 0.08 & -0.32 
\\
$K^{+}_{\mu 3}$ & 2.4 $\pm$ 0.2  & 0.006 $\pm$ 0.16 &  0 $\pm$ 1.0 $^*$  
\\
$K^{0}_{\mu 3}$ &  0  & 0.15 $\pm$ 0.08 &  1.7 $\pm$ 1.0  $^*$  
\\
\hline
\end{tabular}
\end{table}

Each of the four $K_{\ell 3}$ widths is related to $V_{us}$ by a
relation similar to Eq.\ref{masterVud}, modulo an obvious 
change needed to take into account $SU(3)$ breaking effects. 
The $K_{\ell 3}$ widths  thus 
allow us to obtain four determinations~\cite{calc01} 
of $|V_{us}| \cdot f_{+}^{K^0
\pi^-} (0)$ which are independent, up to the small correlations of
theoretical uncertainties from isospin-breaking corrections
$\delta_{\rm EM}$ and $\delta_{SU(2)}$ (almost negligible at present,
see Table~\ref{tab:iso-brk}).  The master formula for a combined
analysis of these modes is:
\begin{equation}
|V_{us}| \cdot f_{+}^{K^0 \pi^-} (0) = \left[ 
\displaystyle\frac{\Gamma_{n [\gamma]}}{ {\cal N}_n \,
 I_n (\lambda, 0) } \right]^{1/2} \, 
\left[ \displaystyle\frac{1}{S_{\rm ew} (M_\rho)} \right]^{1/2} \, 
\displaystyle\frac{1}{1 + \delta^{n}_{SU(2)} + 
 \delta^{n}_{\rm EM} + \frac{1}{2}
\Delta I_{n}(\lambda)}  
\label{masterVus}
\end{equation}
where the index $n$ runs over the four modes 
($n=K^+_{e 3}, K^0_{e3}, K^+_{\mu 3}, K^0_{\mu 3}$).
In writing the above equation, we have chosen 
to ``normalize'' the  form factors for the various modes  to
$f_{+}^{K^0 \pi^-} (0)$, evaluated in absence of electromagnetic
corrections. Differences between the various modes are due to isospin
breaking effects, both of strong ($\delta^{i}_{SU(2)}$) and
electromagnetic ($\delta^{i}_{\rm EM}$) origin, which have been
evaluated at $ {\cal O} (\epsilon p^2)$ in the chiral expansion~\cite{CKNRT}:
\begin{eqnarray}
\tilde{f}_{+}^{n} (0) & = & f_{+}^{K^0 \pi^-} (0) \  
(1 + \delta^{n}_{SU(2)} + \delta^{n}_{\rm EM})~.
\end{eqnarray}

Let us now discuss the theoretical input needed in Eq.\ref{masterVus}.  
In the Table~\ref{tab:iso-brk} we report estimates of the
isospin-breaking parameters~\cite{CKNRT}.
\footnote{~These are based on Ref.~\cite{CKNRT}.
Another calculation~\cite{Bytev} for the $K^+_{e 3}$ mode  
leads to consistent results, though within a different framework.}
The uncertainty on $\delta^{n}_{\rm EM}$ and $\delta^{n}_{SU(2)}$ is due
to incomplete knowledge of the relevant NLO ChPT couplings. 
This is not the dominant uncertainty in the final analysis. 
The phase space corrections refer to the definition of
photon-inclusive width given by Ginsberg~\cite{ginsberg,CKNRT}.  
Although our calculation of phase space corrections for muonic modes
is in progress, in order to include these modes in the
phenomenological analysis, we can use the estimates $\Delta I_{n}, n=
K^{+}_{\mu 3},K^{0}_{\mu 3} $ obtained by Ginsberg~\cite{ginsberg}.
Due to potentially large model uncertainty (e.g. dependence on UV
cutoff) we assign an error bar of $\pm 1 \% $ to these entries.

\begin{table}[t]
\caption{ $K_{\ell 3}$ branching ratios (BR) and slopes 
from PDG fits.
The lifetimes used as input are:
$ \tau_{K^\pm} = (1.2384 \pm 0.0024) \times  10^{-8} \, s $ and
$ \tau_{K_L} = (5.17 \pm 0.04) \times  10^{-8} \, s $. 
\label{tab:PDGinput} }
\vspace{0.4cm}
\centering
\begin{tabular}{|c|c|c|c|}
\hline
Mode            &     BR ($\%$)    &     $\lambda_+$  &   $\lambda_0$  \\
\hline
$K^{+}_{e3}$    &  4.87  $\pm$ 0.06  &  0.0278 $\pm$ 0.0019 &    \\
$K^{0}_{e3}$    &  38.79 $\pm$ 0.27  &  0.0291 $\pm$ 0.0018 &    \\
$K^{+}_{\mu 3}$ &  3.27  $\pm$ 0.06  &  0.033  $\pm$ 0.010  & 0.004 $\pm$
0.009
\\
$K^{0}_{\mu 3}$ &  27.18 $\pm$ 0.25  &  0.033  $\pm$ 0.005  & 0.027 $\pm$
0.006 
\\
\hline
\end{tabular}
\end{table}

The expansion of $f_{+}^{K^0 \pi^-} (0)$ in the quark masses has been
analyzed up to the next-to-next-to-leading order \cite{LeuRo}. At this
level of accuracy we write
\begin{eqnarray}
f_{+}^{K^0 \pi^-} (0) & = & 1 + f_2 + f_4 + {\cal O}(p^6) \ ,   
\end{eqnarray}
where the identity $f_0=1$ follows from current conservation in the
chiral limit. 
Because of the Ademollo-Gatto theorem \cite{ag}, 
local terms are not allowed to contribute to $f_2$, and this 
implies a determination which is practically free of uncertainties:
\begin{equation}
f_2 = -0.023 \; \; .
\end{equation}
As for $f_4$ the situation is much less clear, because low energy constants
(LECs) of the $p^6$ Lagrangian can now contribute.
Various estimates of the size of $f_4$ have been given, and for a 
mini-review we refer to Ref.~\cite{ckm03}.  
In our numerical analysis we use the result of Leutwyler and
Roos~\cite{LeuRo}, based on a model-independent parameterization 
of the asymmetry between kaon and pion wave functions:
\begin{equation}
f_4=-0.016 \pm 0.008 \; \; .  
\label{LRf4}
\end{equation}
We take this result with the understanding that, e.g., 
a value of $f_4$ two sigmas away from the central value 
in Eq.\ref{LRf4} is not strictly forbidden, but rather
unlikely.

From the point of view of the pure chiral expansion, the only
parameter-free prediction which one can make for $f_4$ concerns the
chiral logs. A first step in this direction was made by Bijnens, Ecker
and Colangelo \cite{doublelogs}, who calculated the double chiral logs
contribution to this quantity. The size of this term, however, depends
on the renormalization scale $\mu$. By varying the latter within a
reasonable range, the numerical estimate $|f_4|_{\rm chiral\ logs} \le
0.5 \%$ was obtained.

Recently two groups~\cite{post,bt03} have performed a full calculation
of $f_4$ to $O(p^6)$, which besides the double chiral logs contains
single ones and polynomial contributions. The latter contain LECs of
the $p^6$ Lagrangian, whose value is basically unknown, and make a
numerical estimate difficult.  However, a key observation~\cite{bt03}
is that the relevant local couplings turn out to govern the slope and
curvature of the scalar form factor $f_{0}^{K \pi} (t)$ at $t=0$.
Therefore, at least in principle, such couplings can be related to
observables, through a measurement of the the $t$ dependence of
$f_{0}^{K \pi} (t)$ in $K_{\mu 3}$ decays or through dispersive
analyses of the scalar form factor.
Finally, let us note that lattice calculations of $f_+(0)$  
are certainly called for in order to improve our confidence 
in the range used for $f_4$.

\begin{figure}[t]
\begin{center}
\leavevmode
\begin{picture}(100,180)
\put(10,65){\makebox(50,50){\includegraphics[height= 2 in]{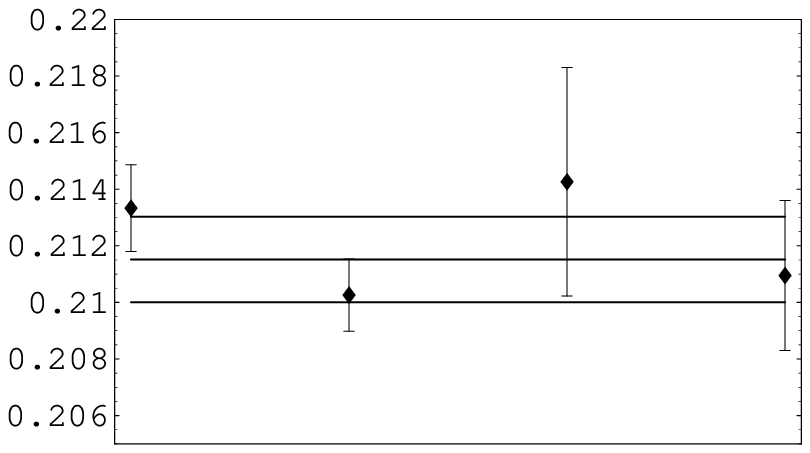}}}
\put(-170,140){\scriptsize{  $|V_{us}| \cdot f_{+}^{K^0 \pi^-} (0)$  }}
\put(-45,10){\scriptsize{$K^+_{e3}$}}
\put(12,10){\scriptsize{$K^0_{e3}$}}
\put(75,10){\scriptsize{$K^+_{\mu 3}$}}
\put(140,10){\scriptsize{$K^0_{\mu 3}$}}
\end{picture}
\caption{ $|V_{us}| \cdot f_{+}^{K^0 \pi^-} (0)$ from the
four $K_{\ell 3}$ modes (and average over the electronic modes).  }
\label{fig:average}
\end{center}
\end{figure} 
%


Using the experimental input reported in Table~\ref{tab:PDGinput}, 
an average over the electronic modes (the muonic modes are irrelevant 
in the average, given the present uncertainties) leads to:
\begin{equation}
|V_{us}| \cdot f_{+}^{K^0 \pi^-} (0)  =  0.2115 \pm 0.0015 \ . 
\end{equation}
The error has been multiplied by a scale factor $S=1.54$, 
as defined by the PDG~\cite{pdg2002}, in order to take 
into account the slight inconsistency between 
$K^{+}_{e3}$ and $K^{0}_{e3}$ results. 
Finally, if we use the Leutwyler-Roos estimate of $f_4$, or
$f_{+}^{K^0 \pi^-} (0)=0.961 \pm 0.008$, we obtain 
\begin{equation}
|V_{us}| =  0.2201 \pm 0.0016_{\rm exp}  \pm 0.0018_{{\rm th}(f_4)} 
=  0.2201 \pm 0.0024  \ .
\label{eq:Vus_fin}
\end{equation}

\section{$V_{ud}$, $V_{us}$,  and CKM unitarity}

Given the tiny size of $V_{ub}$, 
the CKM unitarity constraint of Eq.\ref{eq:unitarity} provides 
a relation between $V_{ud}$ and $V_{us}$. 
Errors on the direct determinations of $V_{ud}$ and $V_{us}$ 
are at the level of 0.1\% and 1\%. Explicitly~\cite{ckm03} one has:
\begin{eqnarray}
|V_{ud}|_{\rm direct} &=&  0.9739 \pm 0.0005 
\qquad \mbox{(Dominated by $n$-$\beta$ and nuclear SFT)}
 \ ; \nonumber \\
|V_{us}|_{\rm direct} &=&  0.2201 \pm 0.0024  
 \qquad \mbox{(Dominated by $K_{e3}$)} \ .
\label{eq:Vus_dir} 
\end{eqnarray}
CKM unitarity implies an independent determination of 
$V_{us}$ (from $V_{ud}$) at the 1\% level,  
\begin{equation}
|V_{us}|_{\rm unit.} = 0.2269 \pm 0.0021 
\end{equation}
to be compared with the direct determination in Eq.\ref{eq:Vus_dir}. 
The $\sim 2 \sigma$ discrepancy between these two determinations could
be attributed to  an underestimate of
i) theoretical uncertainties 
involved in $|V_{ud}|_{\rm direct}$ ; 
ii) theoretical uncertainty in $f_4$ ; 
iii) experimental errors in $K_{\ell 3}$ B.R.s. 
(as probably hinted by the marginal consistency of 
$f_+ (0)  |V_{us}|$ from $K^+_{e3}$ and  $K^0_{e3}$). 
At the moment, in absence of a clear indication of which of the errors
is underestimated, it's probably best to 
treat the two determinations on the same footing, 
and introduce the PDG scale factor in the final error.  
Following this procedure one finds
\begin{equation}
|V_{us}|_{{\rm unit.}+K_{\ell 3}} = 0.2240 \pm 0.0034~.
\label{eq:Vus_global}
\end{equation}

\section{Final remarks}

The pion beta decay ($\pi_{e3}$) provides in principle a unique test
on the existing extractions of $V_{ud}$ from nuclear SFT and neutron
beta decay.  Theory being extremely clean, one has to wait for the
final results from PIBETA~\cite{PSI} to address the impact of this
channel on the determination of $V_{ud}$.

As for $K_{\ell 3}$ decays, our theoretical understanding of radiative
corrections is under control (implying an error on $V_{us}$ of about
0.2 \% ), while improvement is needed in the calculation of 
$SU(3)$ breaking effects.  
The experimental situation will soon  improve thanks to the 
high-statistics measurements of $K_{\ell 3}$ widths 
expected from recently-completed or ongoing experiments,
such as BNL-E865 \cite{E865},  KLOE \cite{KLOE} and NA48 \cite{NA48}. 
It should be  noted how a single present-day measurement allows us 
to extract $|V_{us}|$ at the $\sim 1 \% $ level 
[same as in Eq.\ref{eq:Vus_fin}].      
For example, using the final result from BNL-E865 \cite{E865}, 
${\rm BR}(K^+_{e 3 [\gamma]}) = 
 ( 5.13 \pm 0.02_{\rm stat.} \pm 0.09_{\rm syst.}  
\pm 0.04_{\rm norm.}) \% $, 
and the appropriate phase space correction 
($\Delta I_{K^{+}_{e3}} = (-1.27 + 0.5) \% $),  
one obtains 
\begin{equation}
|V_{us}|_{E865}
 =  0.2272 \pm 0.0030  \ .
\label{eq:Vus_BNL}
\end{equation}
Although this result is in perfect agreement with what is expected
from unitarity, we stress that when combining BNL-E865 with other
existing results one finds:
1) internal inconsistencies within $K^+_{e3}$ measurements;
2) the discrepancy with  $K^0_{e3}$  is worsened;
3) in a combined analysis, the problem with unitarity
persists at the $\sim 2 \sigma$ level. 
In our opinion, a meaningful improvement  will be possible 
once the $K_{\ell 3}$ database is fully updated, including 
not only BNL-E865 but also the results from KLOE and NA48.

\section*{Acknowledgments}

This work has been supported in part by MCYT, Spain (Grant No.
FPA-2001-3031), by ERDF funds from the European Commission, 
and by the EU RTN Network EURIDICE, Grant No. HPRN-CT2002-00311.



\section*{References}
\begin{thebibliography}{99}

\bibitem{ckm03}  M.~Battaglia, A.J.~Buras, P.~Gambino and A.~Stocchi (Eds), 
``The CKM matrix and the unitarity triangle'', 
arXiv:hep-ph/0304132.

\bibitem{ag} 
M.~Ademollo and R.~Gatto,
Phys.\ Rev.\ Lett.\  {\bf 13} (1964) 264.

\bibitem{pdg2002}
K. Hagiwara {\em et al.},  Particle Data Group
Phys. Rev. D {\bf 66} (2002) 010001.

\bibitem{PSI} 
D. Po\v{c}ani\'c (PIBETA Coll.), 
talk given at the CKM Workshop (Durham, April 5-9 2003).

\bibitem{E865}
A.~Sher  et al. (E865 Coll.),
arXiv:hep-ex/0305042.


\bibitem{KLOE}
B. Sciascia (KLOE Coll.), 
talk given at the CKM Workshop (Durham, April 5-9 2003).

\bibitem{NA48} 
D. Madigozhin (NA48 Coll.), 
talk given at the CKM Workshop (Durham, April 5-9 2003).


\bibitem{shortdist}
J.~Erler,
arXiv:hep-ph/0211345 ; 
%
A.~Sirlin,
Nucl.\ Phys.\ B {\bf 196} (1982) 83 ; 
%
W.J. Marciano, A. Sirlin, Phys. Rev. Lett. {\bf 71} (1993) 3629, 
and references therein. 


\bibitem{chpt}  
J.~Gasser and H.~Leutwyler,
Nucl. Phys. {\bf B 250} (1985) 465 ;
Ann. Phys.\  {\bf 158} (1984) 142.

\bibitem{lept}
M. Knecht et al., 
Eur. Phys. J. C {\bf 12} (2000) 469, and references therein. 

\bibitem{CKNRT} 
V.~Cirigliano et al.,  
Eur.\ Phys.\ J.\ C {\bf 23} (2002) 121
[arXiv:hep-ph/0110153]; V. Cirigliano, H. Neufeld, H. Pichl, in progress. 

\bibitem{ginsberg}
E.~S.~Ginsberg,
Phys.\ Rev.\  {\bf 162} (1967) 1570
[Erratum, ibid.\  {\bf 187} (1969) 2280] ; \\
Phys.\ Rev.\  {\bf 171} (1968) 1675
[Erratum, ibid.\  {\bf 174} (1968) 2169] ;  
Phys.\ Rev.\ D {\bf 1} (1970) 229.

\bibitem{cknp02}
V.~Cirigliano et al., 
Eur.\ Phys.\ J.\ C {\bf 27} (2003) 255
[arXiv:hep-ph/0209226].

\bibitem{pibeta}
W. Jaus, Phys. Rev. D {\bf 63} (2001) 053009 ;
A. Sirlin, Rev. Mod. Phys. {\bf 50} (1978) 579.  

\bibitem{calc01}
G.~Calderon and G.~Lopez Castro,
Phys.\ Rev.\ D {\bf 65} (2002) 073032
[arXiv:hep-ph/0111272].

\bibitem{Bytev}
V.~Bytev, E.~Kuraev, A.~Baratt and J.~Thompson,
arXiv:hep-ph/0210049.


\bibitem{LeuRo}
H.~Leutwyler and M.~Roos,
Z.\ Phys.\ C {\bf 25} (1984) 91.


\bibitem{doublelogs}
J.~Bijnens et al.,
Phys.\ Lett.\ B {\bf 441} (1998) 437
[arXiv:hep-ph/9808421].

\bibitem{post}
P.~Post and K.~Schilcher,
arXiv:hep-ph/0112352.

\bibitem{bt03} 
J.~Bijnens and P.~Talavera,
arXiv:hep-ph/0303103.


\end{thebibliography}
\end{document}